\documentclass[amssymb,aps,twocolumn,floats,showpacs]{revtex4}
\usepackage{color,graphicx,pstricks}
\usepackage{natbib}
\usepackage{textcomp}
\usepackage{epstopdf}

\newcommand{\Fig}{Fig.~}
\newcommand{\tj}{$t$-$J$\ }

\begin{document}

\title{Strongly enhanced superconductivity in coupled $t\!\!-\!\!J$ segments}

\author{Sahinur Reja$^1$, Jeroen van den Brink$^1$ and Satoshi Nishimoto$^{1,2}$}

\address{
$^1$Institute for Theoretical Solid State Physics, IFW Dresden, 01171 Dresden, Germany\\
$^2$Institute for Theoretical Physics, TU Dresden, 01069 Dresden, Germany
}

\begin{abstract}
 The \tj Hamiltonian is one of the cornerstones in the theoretical study of strongly correlated copper-oxide based materials. Using the density matrix renormalization group method we calculate the phase diagram of the one-dimensional \tj chain in the presence of a periodic hopping modulation, as a prototype of coupled-segment models.
While in the uniform 1D \tj model near half-filling superconducting (SC) state dominates only at unphysically large values of the exchange
coupling constant $J/t>3$, we show that a small hopping and exchange modulation very strongly reduces the critical coupling to
be as low as $J/t\sim1/3$ -- well within the physical regime. 
The phase diagram as a function of the electron filling also exhibits metallic, insulating line phases 
and regions of phase separation.
We suggest that a SC state is easily stabilized
if \tj segments creating local spin-singlet pairing are coupled to each other -- another example is ladder system.

\end{abstract}

\date{\today} 


\maketitle

\noindent
{\it Introduction ---}
Since its introduction at the end of the 1980's, the \tj model Hamiltonian\cite{zhang_rice} has formed one of the cornerstones in the theoretical study of high temperature superconductors (HTSs). It is a minimal low-energy model for the electronic and magnetic structure of the copper-oxide planes in HTSs and can be derived from the Hubbard model in the strong coupling limit using second order perturbation theory\cite{hubbard_to_tj}. 
Although the initial interest in the \tj model focussed on its two-dimensional (2D) realization, its one-dimensional (1D) version, which is of direct relevance to for instance doped spin-chain materials, provides a number of interesting phases that are also observed in the 2D context, such as a spin gap phase and the occurrence of phase separation\cite{ogata,manmana}. The 1D \tj model actually also displays a superconducting instability, even if all this is in a parameter range where the ratio of the exchange constant $J$ and hopping $t$, near half-filling is $J/t \sim 3$, which is not of relevance to real materials -- the HTSs are rather in the regime where $J/t \sim 1/3$. In the latter regime quasi-1D systems e.g., \tj \ {\it ladders} can support superconductivity \cite{Uehara1996,ladder1,ladder2,ladder3} which is related to the substantial binding energy for two holes in even-leg ladders, giving rise to the presence of preformed Cooper pairs on the rungs. However, in a 1D system this binding energy vanishes for physically relevant values of $J/t$ \cite{ladder4}.  

%
%

\begin{figure}
\begin{center}
\includegraphics[width =.9\columnwidth,angle=0]{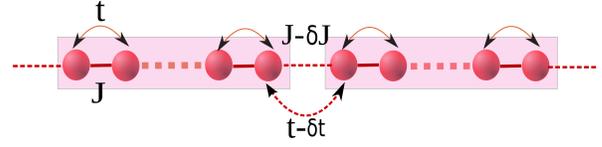} 
\end{center}
\vspace{-.5cm}
\caption{(Color online) Schematic of the model for coupled $t\!\!-\!\!J$ segments. Here 
$t$ and $J$ are the hopping integral and exchange interaction respectively with corresponding
reduction $\delta t$ and $\delta J$ only for the bonds connecting end sites of the segments.}
\label{fig_model}
\end{figure}

In this Letter we consider the 1D \tj Hamiltonian in the presence of a periodic local modulation of  $t$ and $J$: within segments of length $S_l$ the hopping and exchange are constant, but the coupling {\it between} these segments is somewhat weaker, see Fig.~\ref{fig_model}.  In a solid such a local modulation might for instance result from a periodic structural modification or from electronic self-organization.
Using the density matrix renormalization group (DMRG) method we establish that for hole-doped systems close to half filling in the presence of a weak modulation not only a spin-gap forms but also that the holes pair-up already for moderate values of $J/t$. A calculation of the Luttinger parameter shows the formation of a superconducting (SC) state in a large region of the phase diagram, also in the physically relevant low-doping regime with $J/t \sim 1/3$. The transition point weakly depends on segment length and apart from superconductivity also metallic, insulating and phase separated regions are present in the calculated phase diagram.  
  
\noindent
{\it Model and Method ---}
To investigate the effect of coupling of segments forming a 1D chain, we 
consider the model Hamiltonian $H=H_{0}+H_{\delta}$ where $H_{0}$ is the uniform 1D \tj Hamiltonian
$
H_{0} =-t\sum_{ i\sigma}\left ( c^{\dagger}_{i,\sigma} c^{~}_{i+1,\sigma} + H.c. \right ) 
+ J\sum_{i}({\bf S}_{i} \cdot {\bf S}_{i+1}-\frac{1}{4}n_in_{i+1}),
$
where the sum is over all integers $i$ labeling the sites that make up the chain and $H_{\delta}$ describes the local modulation and 
\begin{eqnarray}
H_{\delta}&=&\delta t\sum_{j\sigma}\left ( c^{\dagger}_{j.S_l,\sigma} c^{~}_{j.S_l+1,\sigma} + H.c. \right )\nonumber\\ 
&-& \delta J\sum_{ j }({\bf S}_{j.S_l} \cdot {\bf S}_{j.S_l+1}-\frac{n_{j.S_l}n_{j.S_l+1}}{4}),
\label{model_eq}
\end{eqnarray}
where the sum is over all integers $j$. $H_\delta$ represents the reduction in hopping $\delta t$ and exchange interaction $\delta J$ on the bonds connecting end sites of segments of length $S_l$. 
The modulated bond consist of the rightmost site of $j^{th}$ segment and next site to it, which are indexed by $j.S_l$ and $(j.S_l+1)$ respectively -- a schematic 
of the model is shown in 
\Fig\ref{fig_model}. In the Hamiltonian $c^{\dagger}_{i\sigma}$ is the electron creation operator 
at site $i$ with spin $\sigma$, 
${\bf S_i}$ is the spin-$\frac{1}{2}$ operator, and $n_i$ the electronic number operator. 
%
In order to stay within the perturbative framework in which the  \tj model is derived from the Hubbard Hamiltonian, we retain the direct relation between $\delta t$ and $\delta J$ from second order perturbation theory: as $J={4t^2/U}$, where $U$ is the onsite Coulomb repulsion, we have that  
\begin{eqnarray}
\frac{J-\delta J}{J}=\left( \frac{t-\delta t}{t} \right) ^2
\label{eq:parameters}
\end{eqnarray}
In units of $t$, we are thus left with only two energy scales: $\delta t/t$ and $J/t$. The parameter  ${\delta J}/{J}$ is fixed by the relation in Eq.~[\ref{eq:parameters}].  Clearly for $\delta t=0$ the model reduces to the regular 1D \tj Hamiltonian. The electron density is denoted by $n$.

The quantities of interest e.g., Luttinger parameter, spin gap, binding energy
are calculated by DMRG method\cite{white,dmrg_reveiw} on a lattice with 
upto $288$ sites and finite size extrapolations are performed to obtain them 
in thermodynamic limit (see supplemental material). 
As we have studied the system with segment length $S_l=2,4,6,8,12$
we choose the system size $L=48,96,144,192,240,288$. This allows us to have the
number of electrons $N=nL$ and the number of 
segments in the system to be even so that the ground state corresponds 
to total spin $S_z^T=0$. {All the results are obtained with upto $2000$
basis states, $15$ sweeps and typical discarded weight $\sim 10^{-8}$
leading to error in energy of the same order.}

\noindent
{\it Spin gap and pair binding ---}
We start our discussion with the calculations for the 
spin gap $\Delta_s$ in the system with coupled segments. The usual
\tj model (i.e., $\delta t=0$) in 1D has a spin gap phase, as is found in the studies by
exact diagonalization in small systems\cite{ogata},variational methods\cite{tj_variational},
RG analysis\cite{rg_nakamura} and more recently by DMRG method\cite{manmana}. But 
in all the studies spin gap phase appears only at low {\it electron} density and at large exchange $J/t>2$, a parameter regime that is barely relevant to real materials. 
We show below that the spin gap phase can appear at
small $J/t$ and for weak hole doping (i.e., $n\rightarrow 1$) when a  
non-zero $\delta t$ is introduced.    

\begin{figure}
\begin{center}
\includegraphics[width =1.\columnwidth,angle=-90]{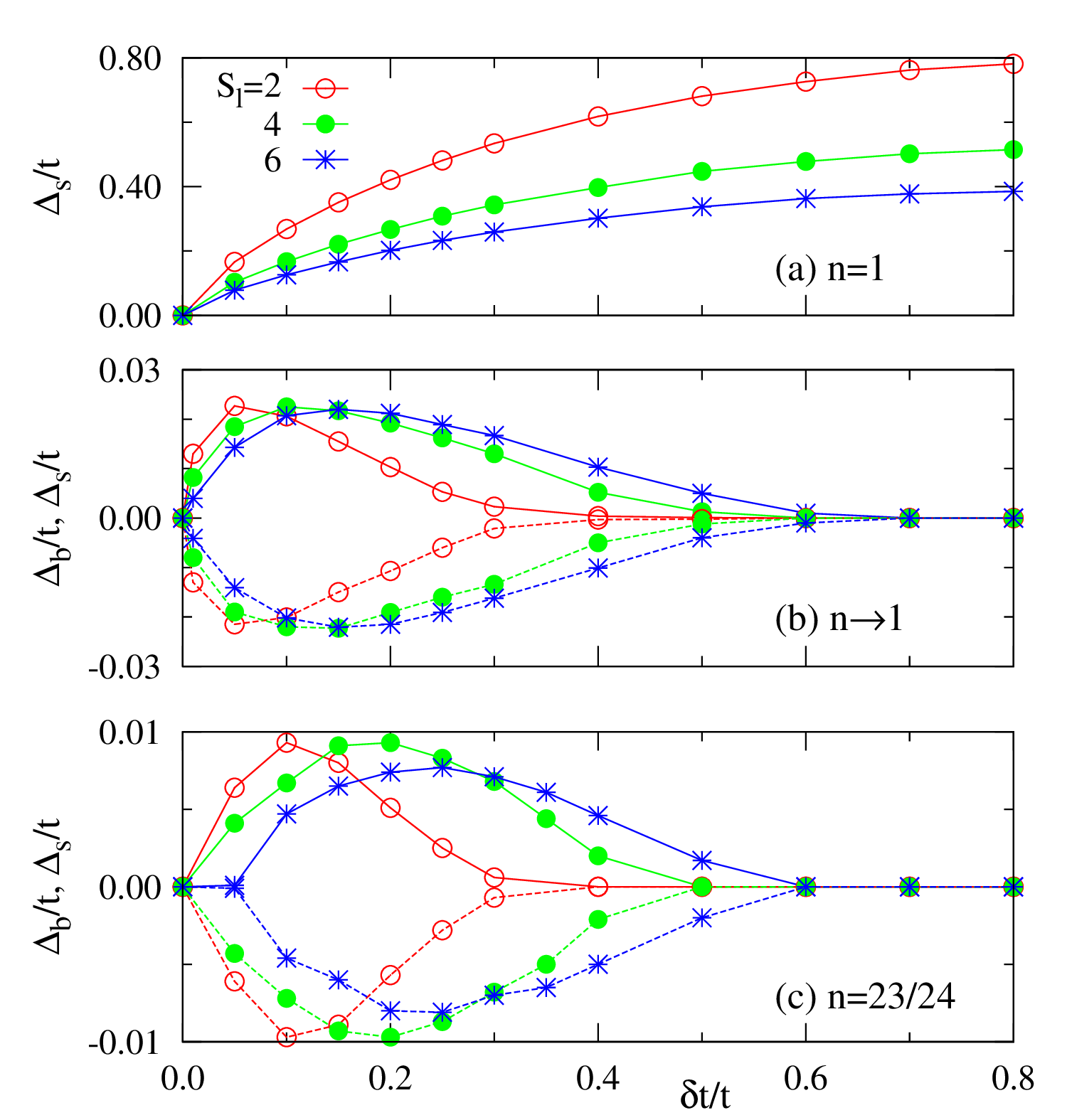} 
\end{center}
\vspace{-.5cm}
\caption{(Color online) The spin gap $\Delta_s$ and binding energy $\Delta_b$ at $J/t=0.8$
as a function of $\delta t$ for (a) half filling $n=1$, (b) in the 
limit of half filling $n\rightarrow 1$ i.e.,
2 holes in the system and 
(c) $n=23/24$. The empty circles (red), filled circles (green) and stared points (blue)
connected with solid lines represent the spin gap for segment size $S_l=2$, $4$ and $6$
respectively. The corresponding circles and points connected with dotted lines 
represent binding energy.}
\label{fig_delta_s_be}
\end{figure}

The singlet-triplet excitation energy is given by the energy difference
$\Delta_s=E(N,S_z^T=1)-E(N,S_z^T=0)$. Here $E(N,S_z^T)$ is the ground state energy
with quantum numbers $N$ and total $z$-component of spins $S_z^T$.
At half-filling, $n=1$, the uniform \tj model corresponds to a Mott insulator
with zero spin gap. But if  $\delta t$ is switched on, a spin gap $\Delta_s$ forms and increases monotonically, see \Fig\ref{fig_delta_s_be}(a). The gap reaches its maximum in the limit $\delta t /t =1$, where it corresponds to the finite size gap of decoupled Heisenberg segments.
However, the situation changes drastically when holes are doped into the system, see \Fig\ref{fig_delta_s_be}(b) and (c). The overall spin-gap is much reduced but kept to be finite, and reaches its maximum for {\it small} $\delta t/t$ and gradually vanishes for larger values of the hopping modulation.
%
%
\Fig\ref{fig_delta_s_be}(b) shows the spin gap $\Delta_s$ for $J/t=0.8$ 
at different $\delta t$ in the limiting density $n\rightarrow 1$ which corresponds to putting only two holes in different system sizes and extrapolating 
$\Delta_s$ to thermodynamic limit (see supplemental material). The calculations were done for segment sizes $S_l=2, \ 4$ and $6$. 
Surprisingly, a very small $\delta t=0.01$ is already enough to produce a sizable spin gap in the system. The spin gap increases first as in half-filled case, but then decreases to zero for larger $\delta t$. {It is interesting that the maximum of the spin gap
is not really sensitive to the segment size. We notice that the occurrence of maximum of $\Delta_s$ shifts to  higher $\delta t$ if we increase segment size $S_l$. It is related to slower reduction of (electron or hole) bandwidth by $\delta t$ for larger $S_l$.}
Also shown in the figures is  the pair binding energy, which measures the stability of pairing and is defined as 
$\Delta_b=E(N\pm 2,S_z^T=0)+E(N,S_z^T=0)-2E(N\pm 1,S_z^T=\pm 1/2)$.
The plot for $\Delta_b$ is shown in \Fig\ref{fig_delta_s_be}(b) for $S_l=2, \ 4$ and $6$ 
with the same points and colors as $\Delta_s$ but connected by dotted lines. {The negative 
values of $\Delta_b$ indicates the stability of pairing and  a relation $\Delta_s=|\Delta_b|$ in a metallic (M) regime
suggests an occurrence of spin-singlet SC state.}
The spin gap and pair binding energy at a finite hole density of  $n=23/24$ are shown in \Fig\ref{fig_delta_s_be}(c). 
{Compared to the limit of half filling $n\rightarrow 1$, their magnitudes are smaller but the over-all behaviors are
almost unchanged.}


\begin{figure*}
\begin{center}
\includegraphics[width =.9\textwidth,angle=0]{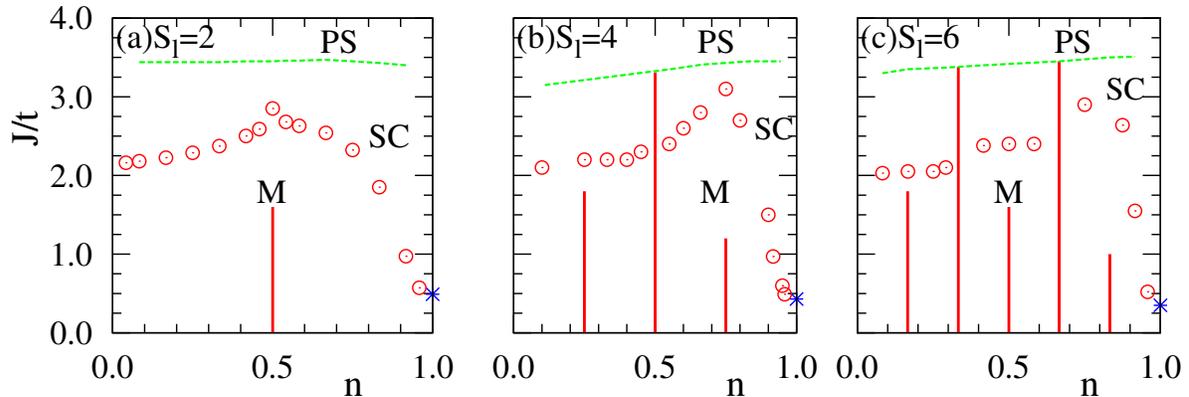} 
\end{center}
\caption{(Color online) (a)-(c) Phase diagrams on $n\!\!-\!\!J/t$ 
plane at $\delta t/t=0.2$ for $S_l=2$, $4$ and $6$ systems 
respectively containing different phases: Superconducting (SC),
Metalic (M) and Phase separated (PS) as indicated. The points (empty circles)
represent M to SC transition and the point (star) indicates density $n\rightarrow 1$
 (system with 2 holes).
The vertical lines at different commensurate fillings represent insulating line-phases.}
\label{fig_phase_diag}
\end{figure*}

\noindent
{\it Superconductivity and Phase diagram---}
The DMRG calculations allow us to map out the phase diagram for the modulated 1D \tj Hamiltonian including also potential states.
It should be noted that the uniform \tj model shows a transition from M to 
SC state at $J/t\geq 2$ in low electron density regime. An even higher $J/t$
is necessary to get SC phase with increasing electron density\cite{manmana}.
In order to examine the possibility of superconductivity in system with 
coupled segments at different 
electronic density $n$, we determine the metal-SC transition by calculating the Luttinger parameter: $K_\rho<1$
for a paramagnetic metal and $K_\rho>1$ for a superconductor. In the thermodynamic limit the Luttinger
parameter is determined from the slope of the structure factor for the 
density-density correlation at wave vector $k\rightarrow 0$ 
limit\cite{giamarchi,rt_clay}. 
To avoid difficulties in the calculation of real-space density-density correlations at large distances and the subsequent Fourier transform at small $k$,
we calculate structure factor directly in momentum space\cite{satoshi1} and extract $K_\rho$ in thermodynamic limit by finite size scaling (see supplemental material).
This requires calculating $K_\rho$ for different system sizes (upto $288$ sites). 
%


We determined the crossing point $K_\rho=1$  for fixed different densities $n$, segment lengths $S_l$, and modulation
strengths $\delta t$, that separates the M and SC phases as a function of $J/t$. The critical value is denoted by $J^c/t$ hereafter.
At very large $J$, the system separates into electron- and hole-rich region and the onset of phase separation (PS) has been
determined by inverse compressibility (see supplemental material).
 The resulting $n\!\!-\!\!J/t$ phase diagrams containing PS, M and SC phases are shown in \Fig\ref{fig_phase_diag}(a)-(c) for $\delta t/t =0.2$ and $S_l=2$, $4$ and $6$, respectively. 
{The (blue) star symbol represents the critical point $J^c/t$ in $n\rightarrow 1$ limit. It is remarkable that
close to half filling (even for a system with only two holes i.e., $n\rightarrow 1$)
$J^c/t$ is
significantly reduced for all the cases of three $S_l$'s as shown in \Fig\ref{fig_phase_diag}.
Apparently, the critical value can be as small as $J^c/t\sim0.35$ for a relatively weak modulation $\delta t/t=0.2$.}



\begin{figure}
\begin{center}
\includegraphics[width =.7\columnwidth,angle=-90]{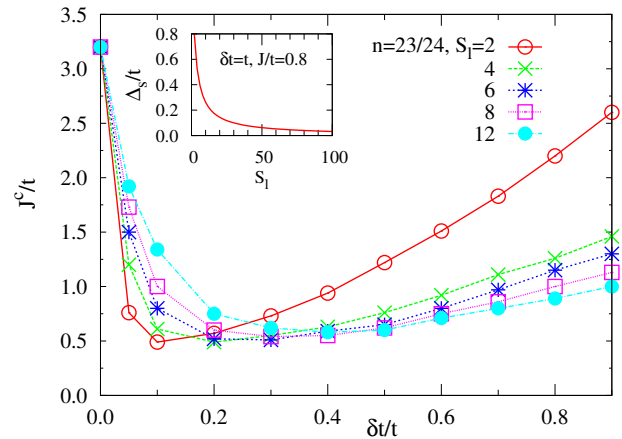} 
\end{center}
\vspace{-.5cm}
\caption{(Color online) This shows the transition from metallic to superconducting
phase at density $n=23/24$ as a function $t_2$ and for systems with segment sizes 
$2, 4, 6, 8, 12$ as indicated. In the inset we plot $\Delta_s$ for systems with decoupled 
segments of different sizes at $J/t=0.8$. } 
\label{fig_diff_bs}
\end{figure}

{This observation motivates to study how the critical point $J^c/t$ changes as a function of $\delta t/t$.
Since a large reduction of $J^c/t$ can be acquired near half filling, we fix the filling at $n=23/24$ and estimate
the critical point.
The critical value $J^c/t$ at $n=23/24$ is shown in \Fig\ref{fig_diff_bs} for $S_l=2, 4, 6, 8, 12$.
For each segment length even a weak modulation $\delta t$ causes a large drop in $J^c/t$ from $\sim3.2$
at the uniform case $\delta t=0$.} Depending on the segment size the lowest $J_c/t$ is reached for $\delta t/t$ between
$\sim 0.1$ and $\sim 0.3$.
The calculations imply that an increase of $\delta t$ helps electrons in each segment to form pairs which can move easily unless $\delta t/t$ becomes too large, which then tends to localize electrons on individual segments for large $\delta t/t$. This explains
the steep minimum and then slow rise in $J^c/t$.
At small $\delta t/t$, the transition point $J^c/t$ increases with $S_l$ because the pair formation gets weakened as
the number of sites in a segment becomes larger. Ultimately, at very large $S_l$, this situation will correspond to
the uniform \tj model.
{On the other hand at large $\delta t/t$, the pair formation is most effective but the movement of pairs 
is restricted due to the narrowing of bandwidth and a high $J^c/t$ is given accordingly.
An intriguing thing is that for $S_l=12$ the strong reduction of $J^c/t$ is obtained despite
$J^c/t\sim3.2$ for any $\delta t/t$ in the $S_l \to \infty$ limit. It means that a system with $S_l=12$ is still far from
the uniform \tj model.}
We also notice a sharp decrease in $J^c/t$ as we increase $S_l=2$ to $12$ initially at large $\delta t/t$ and then almost
saturates. To explain this behavior, we calculate the spin gap of decoupled segments (i.e., $\delta t/t=1$) of 
different sizes starting from $S_l=2$. As expected, we see a sharp decrease of spin gap with $S_l$ as shown in the inset
of \Fig\ref{fig_diff_bs}. 
{Plots in \Fig\ref{fig_diff_bs} also explains the appearance of 
finite spin gap discussed in \Fig\ref{fig_delta_s_be}(c)
for density $n=23/24$ at $J/t=0.8$. The critical $J^c/t$ where the M to SC transition
occurs is lower than $J/t=0.8$ for some range of $\delta t/t$ and for different $S_l$.
This means the system develops a finite spin gap entering into SC phase.}

\noindent
{\it Insulating line phases ---}
Unlike the uniform \tj model, we notice some insulating regions (vertical lines 
in \Fig\ref{fig_phase_diag}) appearing at 
various commensurate fillings. This type of insulating region 
at particular commensurate fillings has been discussed in 
ladder systems\cite{insul_ladder}. Due to the restriction in hopping as 
we introduce $\delta t$, the electrons may form super structure in each segment
which leads to insulating behavior of the system. The insulating to M
transition at $n=1/2$ which is shown in 
\Fig\ref{fig_phase_diag}(a) for $S_l=2$ system
has been discussed in terms of melting 
of Mott insulating states\cite{satoshi2}. 
We briefly recap its intuitive understanding
which will guide us to understand the insulating regions at other fillings
and $S_l=4, 6$ systems. 

Let us first consider $S_l=2$ system with no exchange interaction 
i.e., $J\rightarrow 0$ in 
$U\rightarrow\infty$ limit. Then the tight binding model 
with $\delta t>0$ at $n=1/2$ forms bonding and 
antibonding states with spin $\sigma$ i.e., $t-$dimer states:  
$|t\!-\!dim\rangle_\sigma=(|\sigma \;0\rangle\pm |0\; \sigma\rangle)/\sqrt{2}$ 
separated by a gap $\Delta_d=2\delta t$ at each segment.
Due to this $t-$dimerization one electron per segment is in the
bonding state at this commensurate density of
$n=1/2$ and the system becomes a Mott insulator.         
But if we increase $J$, a tendency to make singlet pairs is introduced 
at each segment due to $J$-dimerization which 
leads to a transition to M states. If we increase $J$ further where
$J$-dimerization dominates, the system with these singlet pairs becomes SC
and finally goes to PS at large $J$. Such a melting of insulating states happen also in 
$S_l=4$ system at $n=1/4, 3/4$ and $S_l=6$ system at $n=1/6, 3/6, 5/6$
as shown in \Fig\ref{fig_phase_diag}(b) and (c), respectively. No effective pairing with odd 
number of electrons in a segment at these electronic fillings causes the melting of insulating
states with increasing $J$. However, even number of electrons in a segment as in $S_l=4$
system at $n=1/2$ and $S_l=6$ system at $n=2/6, 4/6$ prevent such melting of insulating states
by making effective pairing between themselves even with increasing $J$. So the 
insulating states continue to exist until the PS boundary at these fillings with increasing $J$
as shown in \Fig\ref{fig_phase_diag}(b) and (c).
These insulating line phases and metal-insulator transitions 
are captured by calculating charge gap 
defined as $\Delta_c=(E(N+2)+E(N-2)-2E(N))/2$ for different values of $J$. 

\noindent
{\it Conclusions}:
We have shown using density matrix renormalization group method that a metal to superconducting transition can occur in
a physically relevant regime of exchange interaction and hopping (i.e.,$J/t\sim1/3$) in 1D system with coupled \tj segments.
A moderate modulation $\delta t/t$ in the hopping causes a superconducting state for electron densities close to half filling.
As compared to uniform 1D \tj model, this implies an order of magnitude reduction in critical exchange interaction to hopping ratio.
%
We have presented full $n\!\!-\!\!J$ phase diagrams at $\delta t/t=0.2$ with different segment sizes containing besides phase separation, metallic and superconducting phases also insulating line phases at commensurate doping concentrations. 
These results may be of relevance to certain quasi 1D materials, including cuprates in which typical 
exchange coupling is $J\sim t/3$\cite{cuprates_J}. An example is the spin-Peierls system 
CuGeO$_3$\cite{spin_peierls} which consists of linear spin-$\frac{1}{2}$ CuO$_2$ chains 
with alternating exchange coupling and hopping strength, corresponding to a $S_l=2$ system in our language. The results above suggest that doping with few holes may turn the system to be superconducting. Such values of exchange coupling and hopping strengths can in principle also be achieved in 1D ultracold fermionic quantum gases\cite{ultracold1} and with polar molecules\cite{ultracold2} on optical lattices. It will be most interesting to establish whether in such quantum simulator experiments a superconducting state can be stabilized by a relatively benign modulation of the 1D \tj Hamiltonian as we propose here.  
{In this Letter, the segments are open chain coupled linearly as a simplest 1D case. However, 
any shape of segment can be allowed if it creates locally finite spin excitation gap. The network between the segments
would be also flexible. We think that this simple condition possibly makes a great help for superconducting material design.
For example, the ladder \tj system is a special case of the coupled-segment systems.
}

\noindent
{\it Acknowledgments}: This work is supported by SFB 1143 of the Deutsche Forschungsgemeinschaft.



\end{document}


\title{Supplementary Material for\\ ``Strongly enhanced superconductivity in coupled $t\!\!-\!\!J$ segments''}

\author{Sahinur Reja$^1$, Jeroen van den Brink$^1$ and Satoshi Nishimoto$^{1,2}$}
\address{
$^1$Institute for Theoretical Solid State Physics, IFW Dresden, 01171 Dresden, Germany\\
$^2$Institute for Theoretical Physics, TU Dresden, 01069 Dresden, Germany
}

\date{\today}
\begin{abstract}
 Here we present a detailed plots of finite size scaling analysis to get spin gap, charge gap and 
 Luttinger parameter in thermodynamic limit. Luttinger parameter has been used to determine the 
 transition between metallic to superconducting phases whereas spin gap and binding energy indicate  
 the stability of pairing of electrons. Insulating line phases are captured by finite charge gap
 and onset of phase separation is obtained from inverse compressibility.    
\end{abstract}

\maketitle

\begin{figure}
\subfloat[]{\includegraphics[width = .33\textwidth,angle=-90]{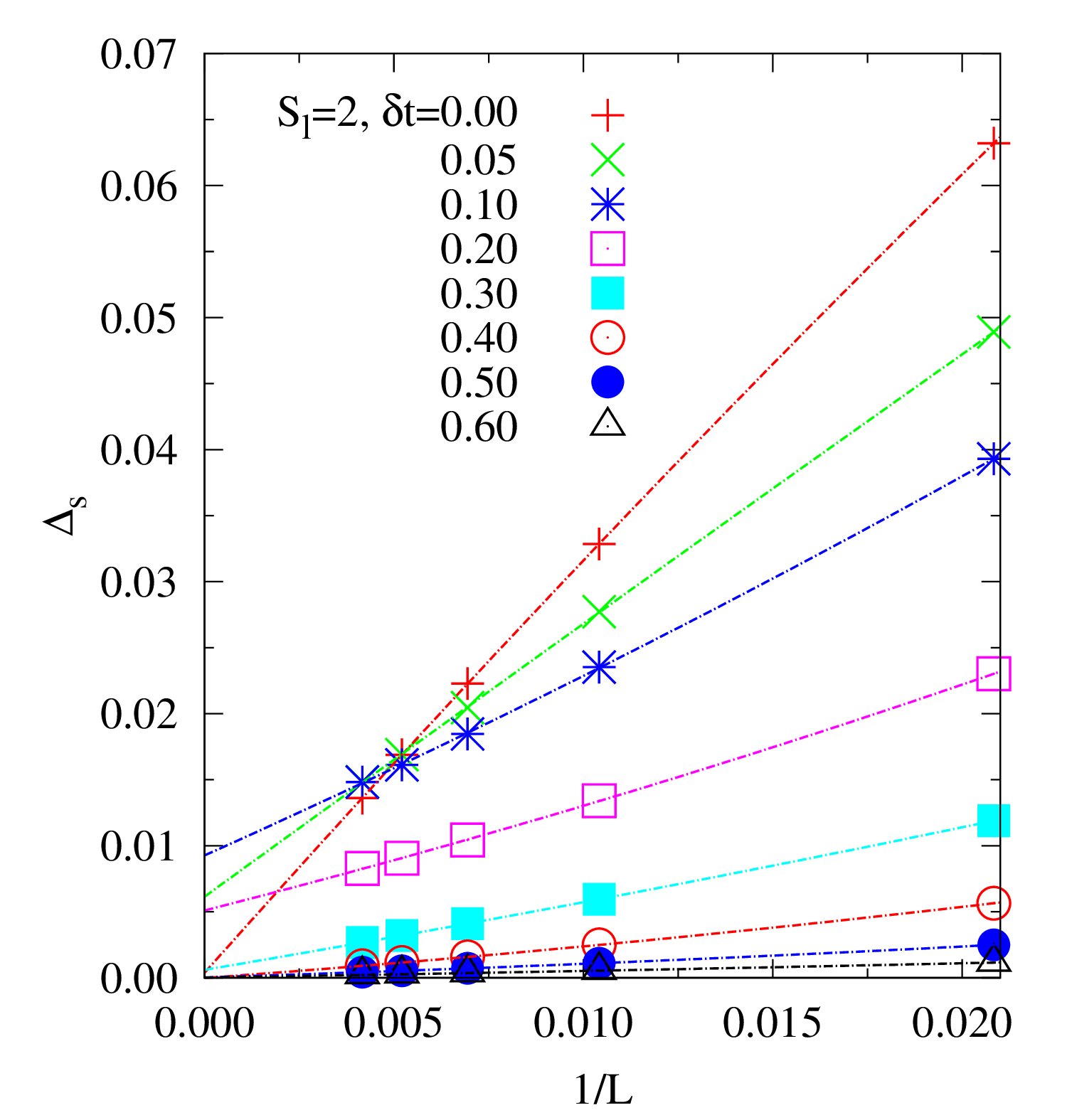}} 
\subfloat[]{\includegraphics[width = .33\textwidth,angle=-90]{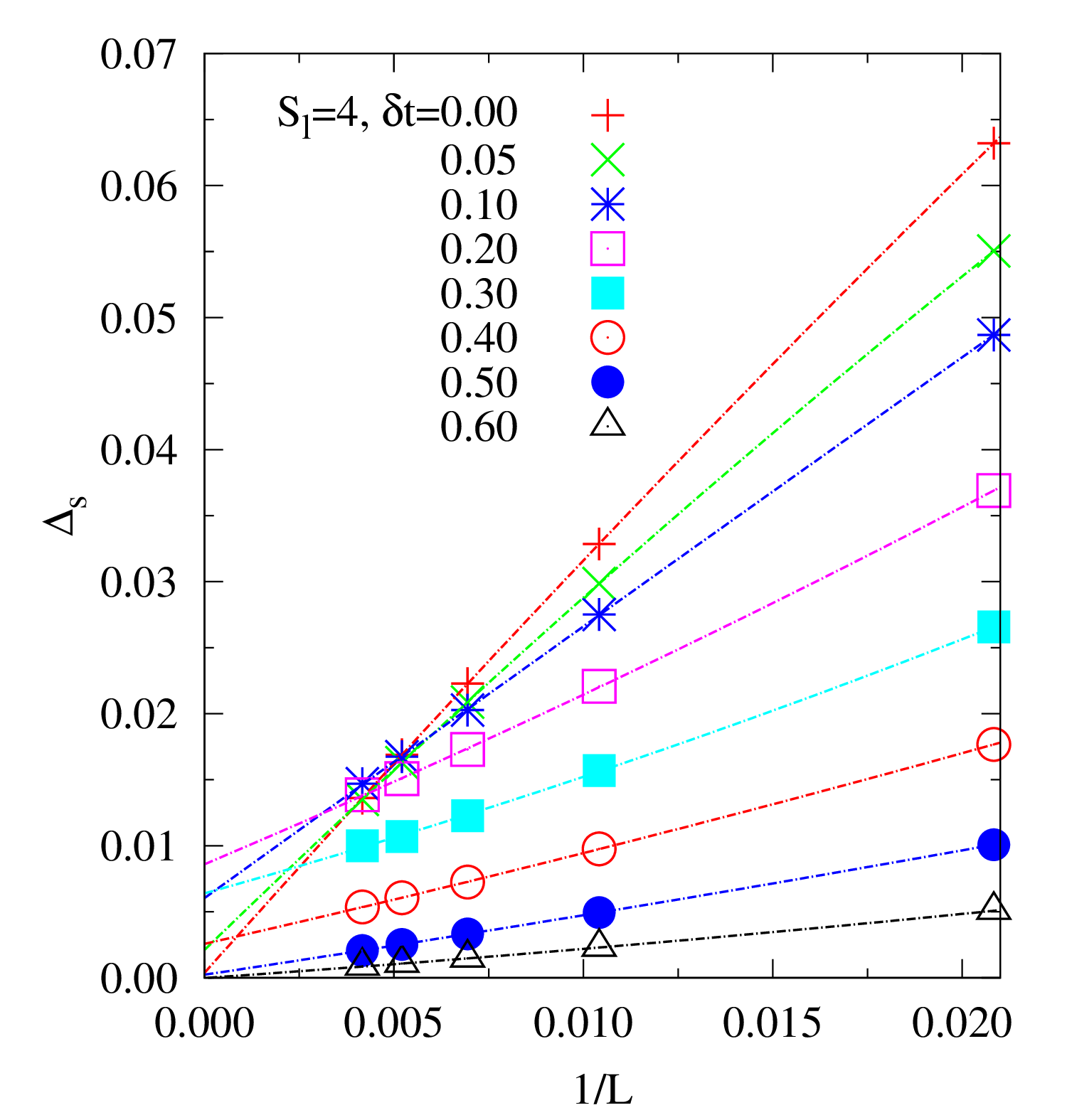}}
\subfloat[]{\includegraphics[width = .33\textwidth,angle=-90]{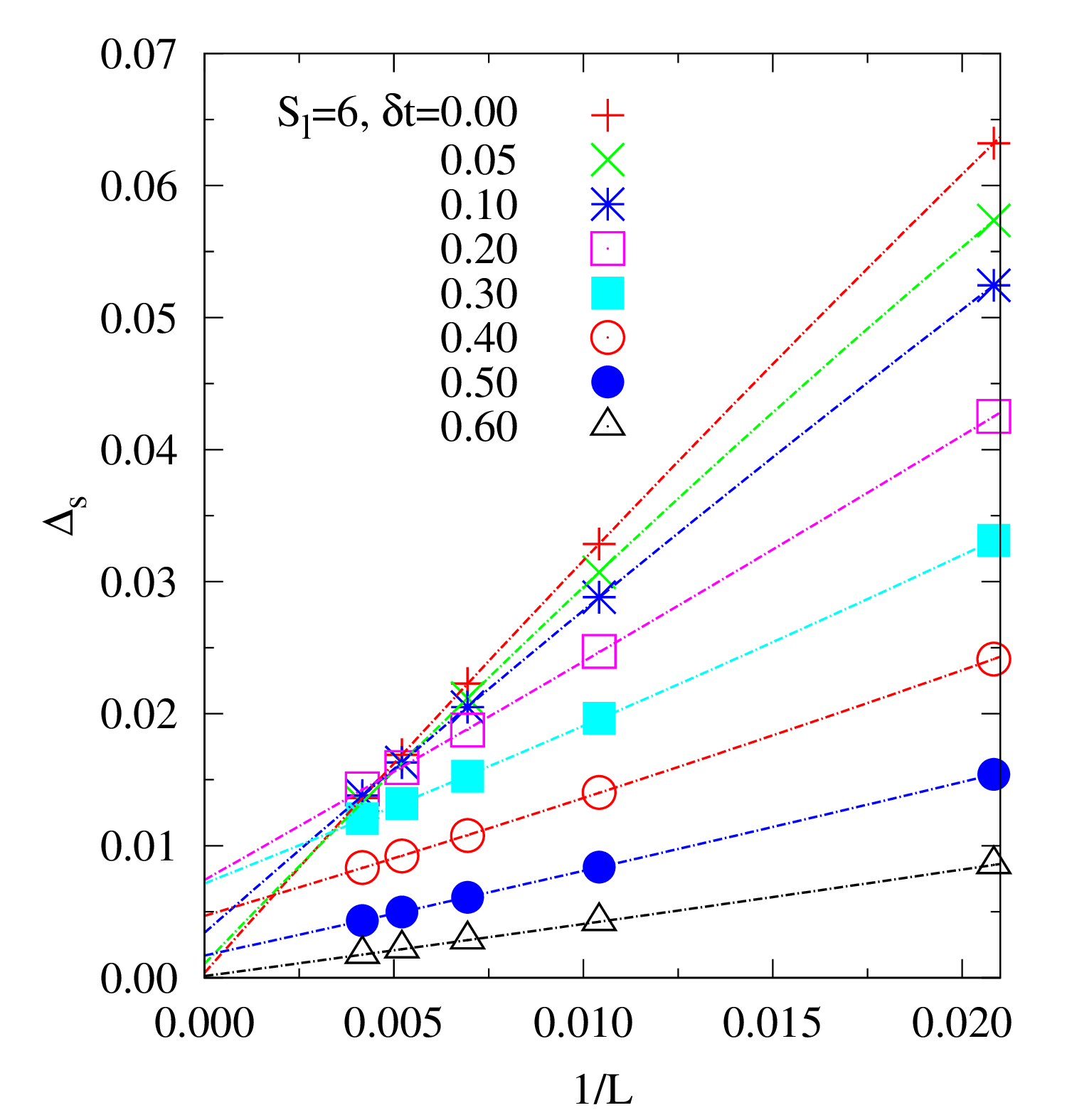}} 
\caption{(Color online) Spin gap extrapolation to thermodynamic limit at finite electronic filling $n=23/24$.
These plots are for different hopping modulation $\delta t$ and segment size $S_l=2, 4, 6$ in 
(a), (b) and (c) respectively with exchange coupling is $J/t=0.8$}
\label{dels}
\end{figure}

\begin{figure}
\subfloat[]{\includegraphics[width = .33\textwidth,angle=-90]{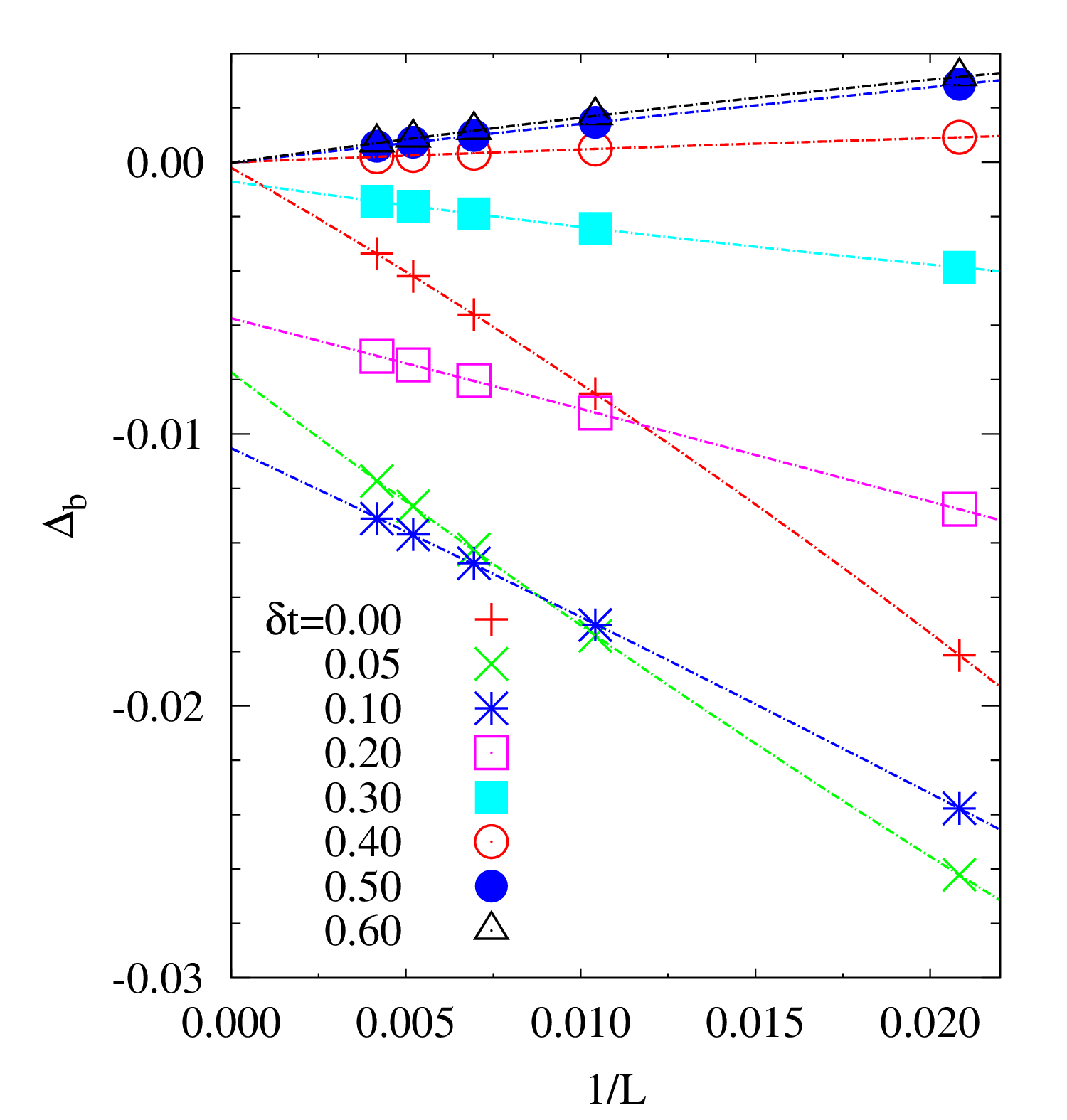}} 
\subfloat[]{\includegraphics[width = .33\textwidth,angle=-90]{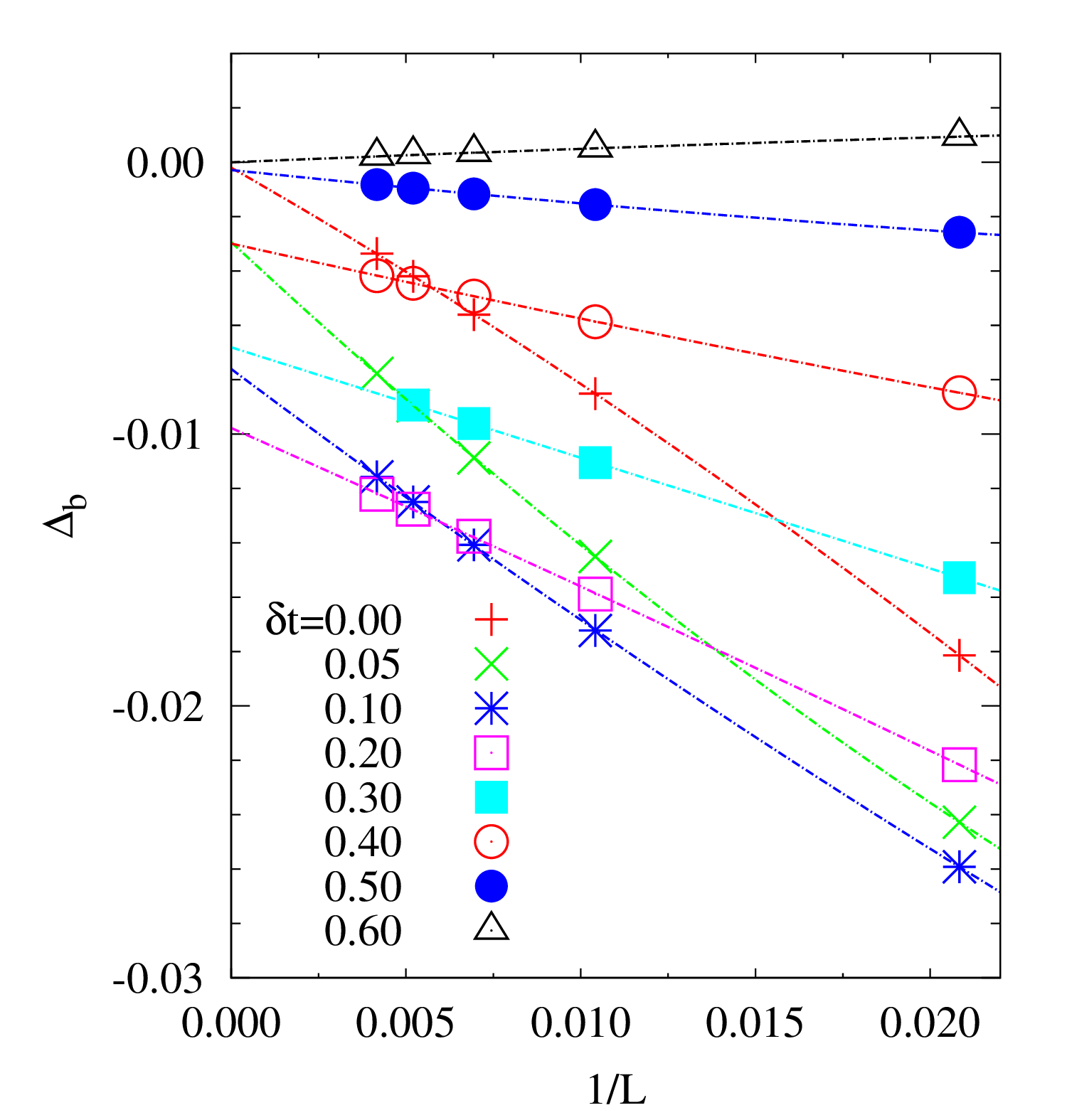}}
\subfloat[]{\includegraphics[width = .33\textwidth,angle=-90]{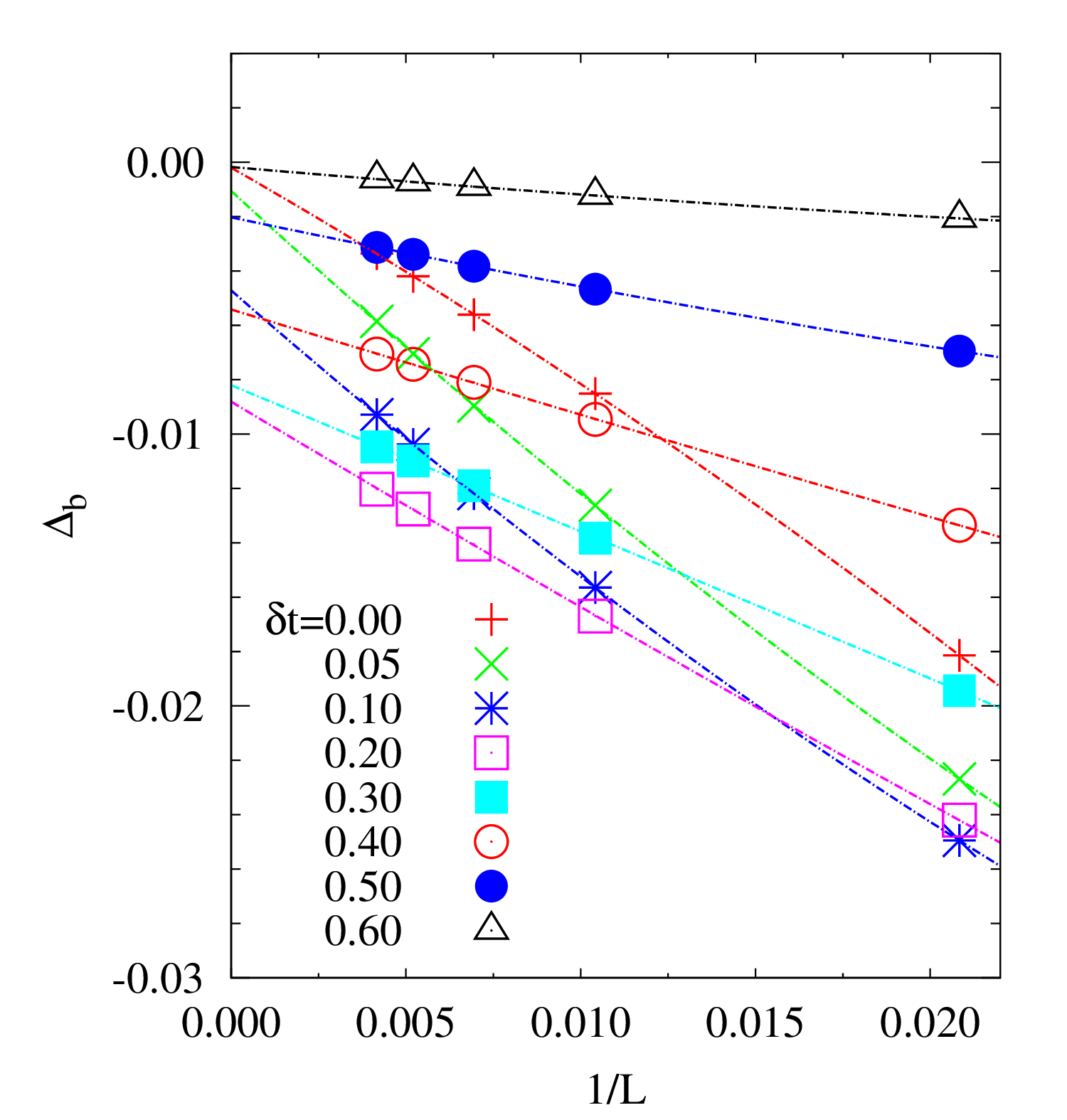}} 
\caption{(Color online) Binding energy extrapolation to thermodynamic limit at finite electronic filling $n=23/24$.
These plots are for different hopping modulation $\delta t$ and segment size $S_l=2, 4, 6$ 
in (a), (b) and (c) respectively with exchange coupling is $J/t=0.8$}
\label{delb}
\end{figure}

We use DMRG method\cite{white,dmrg_reveiw} to study the system consisting of coupled 
\tj segments in presence of periodic hopping modulations on a lattice with 
total upto $288$ sites and different segment size $S_l$. We calculate each of 
the quantities of interest e.g., 
Luttinger parameter, spin gap, binding energy, charge gap  
in thermodynamic limit by finite size extrapolation.  
As we have studied the system with segment length $S_l=2,4,6,8,12$
we choose the system size $L=48,96,144,192,240,288$
and more in some calculations. This allows us to have the
number of electrons $N=nL$ and the number of 
segments in the system to be even so that the ground state corresponds 
to total spin $S_z^T=0$. All the results are obtained with at least $500$
basis states, $15$ sweeps and typical discarded weight $\sim 10^{-8}$
leading to error in energy of the same order.

We make use of Luttinger parameter $K_\rho$ to find the metallic ($K_\rho<1$) to 
superconducting ($K_\rho>1$) transition as a function of exchange coupling 
$J$ and electronic density $n$. This enables us to map out the $n-J$
phase diagrams presented for $S_l=2, 4, 6$ systems. The spin gap and binding energy indicate the 
stability of electron pairing in superconducting state. Insulating line phases at different
commensurate fillings are captured by charge gap calculation. Finally inverse compressibility
has been used to estimate the phase separation boundary.

\section{Spin gap and pair binding}

Spin gap represents singlet to triplet excitation energy and is defined as:
\begin{eqnarray}
\Delta_s=E(N,S_z^T=1)-E(N,S_z^T=0)
\end{eqnarray}
where $E(N,S_z^T)$ is the ground state energy
with quantum numbers $N=nL$ and total $z$-component of spins $S_z^T$.
\Fig\ref{dels} shows the finite size scaling of spin gap.
We choose finite electronic density
$n=23/24$ and exchange coupling $J/t=0.8$ for different segment size 
$S_l=2, 4$ and $6$ as indicated. In each case, the usual \tj model i.e., hopping modulation 
$\delta t=0$ gives zero spin gap.
As we introduce finite  $\delta t$, the system develops finite spin gap and vanishes 
again at higher $\delta t$. Similarly corresponding binding energy: 
$\Delta_b=E(N\pm 2,S_z^T=0)+E(N,S_z^T=0)-2E(N\pm 1,S_z^T=\pm 1/2)$ extrapolations are 
shown in \Fig\ref{delb}. The spin gap and binding energies are found to be almost equal.

\section{Luttinger Parameter}
We find the transition
from metallic to SC phase by calculating the Luttinger parameter $K_\rho<1$
for metallic and $K_\rho>1$ for superconducting phase. In thermodynamic limit Luttinger
parameter is calculated from the slope of the structure factor for the 
density-density correlation at wave vector $q\rightarrow 0$ 
limit\cite{giamarchi,rt_clay}.
\begin{eqnarray}
K_\rho=\pi \lim_{q\rightarrow 0^+}\frac{\widetilde{W}(q)}{q} 
\label{K_rho_eq}
\end{eqnarray}   
 where for finite systems in numerical simulations $q=2\pi/L$, $L$ being the system size and 
 the Fourier transform of correlation, 
 \begin{eqnarray}
\widetilde{W}(q)=\frac{1}{L}\sum_{l=1}^{L}(\langle n_in_{i+l}\rangle-
 \langle n_i\rangle\langle n_{i+l}\rangle)e^{-iql}\nonumber
 \end{eqnarray}

The accuracy of $K_\rho$ suffers in this approach due to the difficulties in accurate 
calculation of density-density correlations, specially at large distances and subsequently
the Fourier transform at small $q$. So we calculate structure factor directly
in momentum space\cite{satoshi1} and extract $K_\rho$ in thermodynamic limit by
finite size scaling.
 
 For this we define an operator in momentum space:
 \begin{eqnarray}
\widetilde{N}(q)=\frac{1}{L}\langle\Psi_0|\tilde{n}(q)\tilde{n}(-q)|\Psi_0\rangle
 \end{eqnarray}

where $\tilde{n}(q)=\sum\limits_{l,\sigma}c^{\dagger}_{l\sigma} c^{~}_{l\sigma}e^{-iq(l-r_c)}$;
$r_c=(L+1)/2$ being the middle position of the chain. Note that $\widetilde{N}(q)$ and 
$\widetilde{W}(q)$ are different, but becomes identical only in thermodynamic limit. So 
finally we get $K_\rho$ by targeting
not only the ground state $\Psi_0$ but also the state $\tilde{n}(q)|\Psi_0\rangle$ as:
\begin{eqnarray}
K_\rho=\lim_{L\rightarrow\infty}\frac{L}{2}\widetilde{N}(2\pi/L)
\end{eqnarray}

\begin{figure}
\subfloat[]{\includegraphics[width = .33\textwidth,angle=-90]{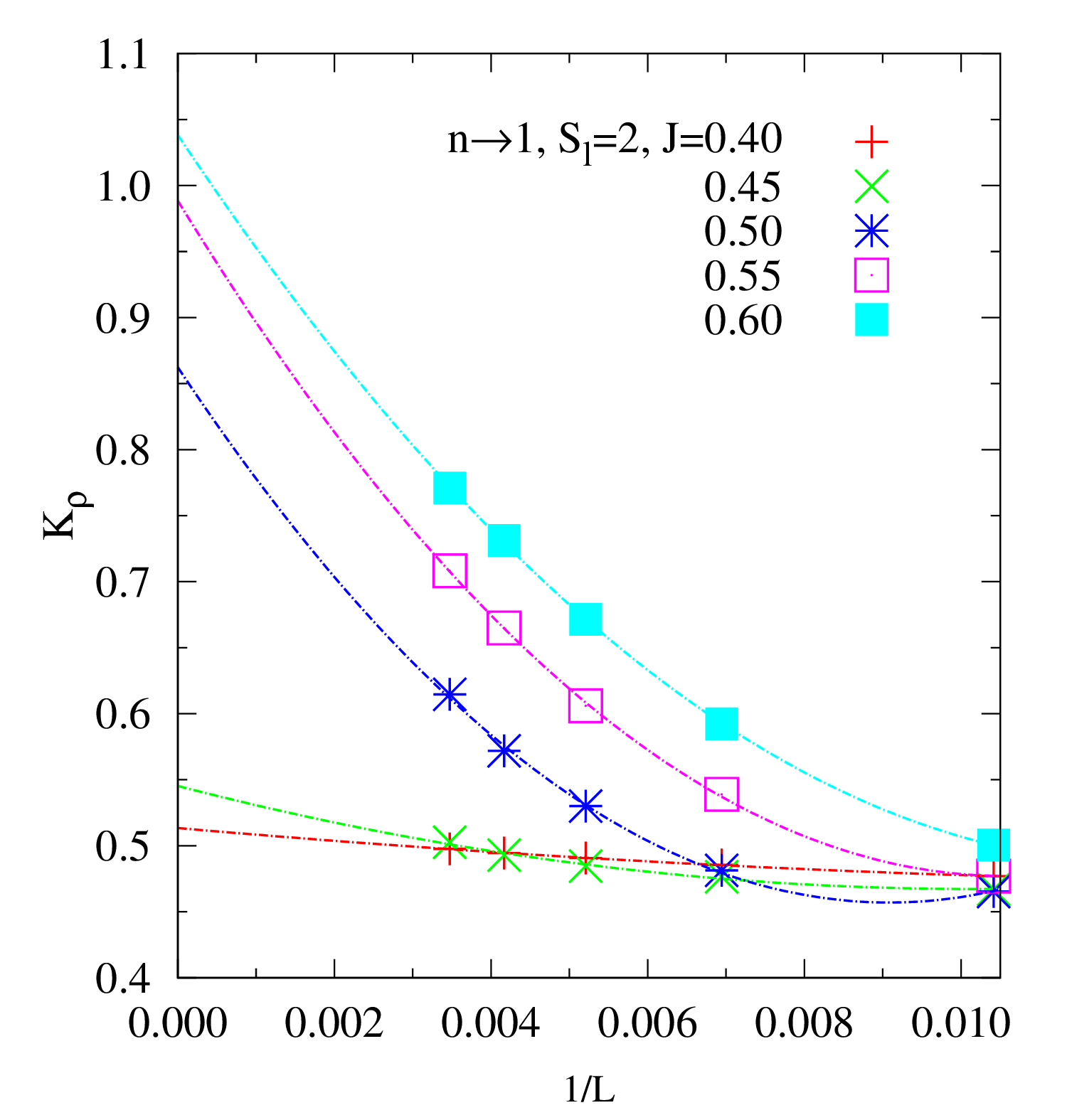}} 
\subfloat[]{\includegraphics[width = .33\textwidth,angle=-90]{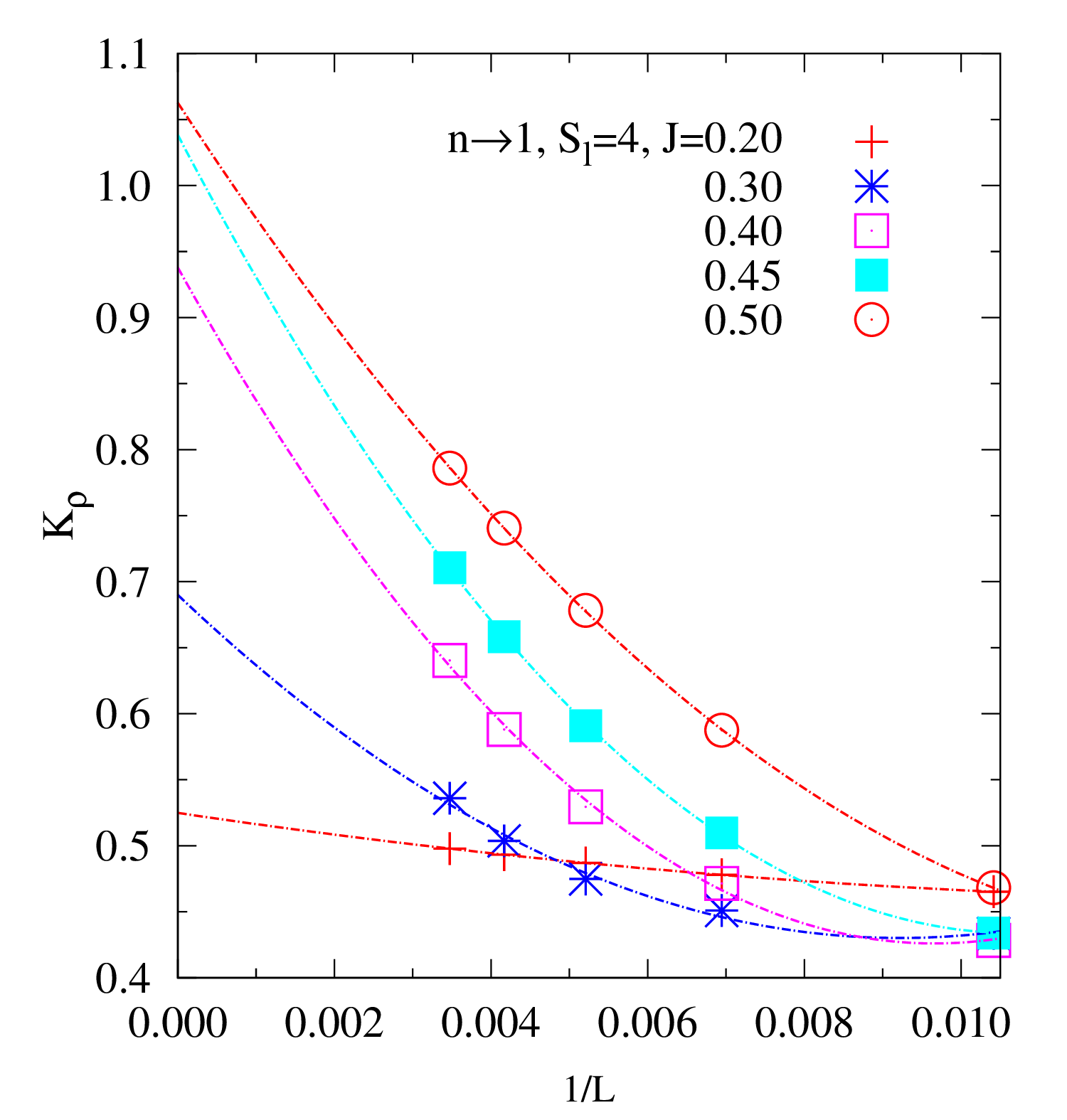}}
\subfloat[]{\includegraphics[width = .33\textwidth,angle=-90]{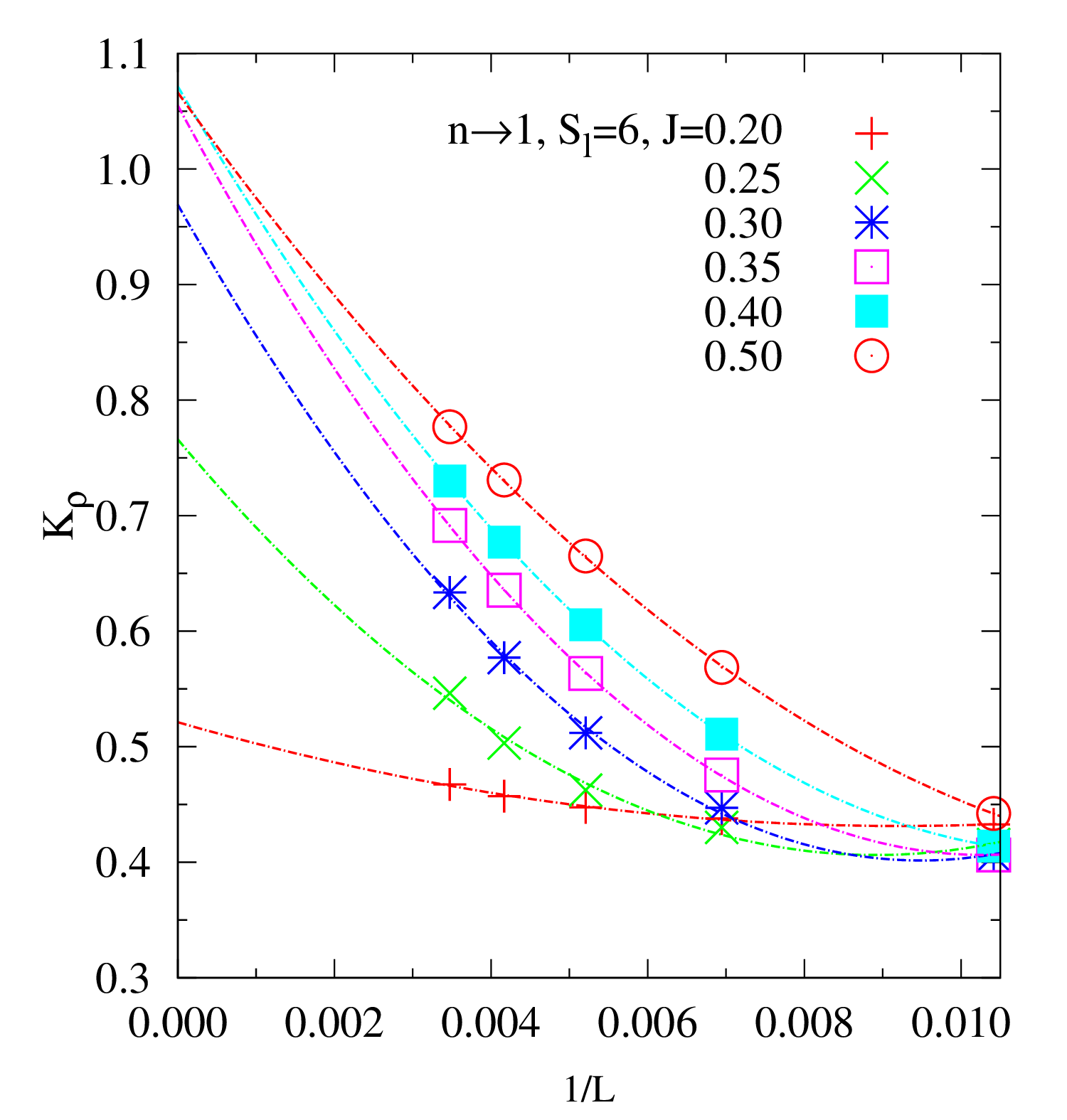}} 
\caption{(Color online) The Luttinger parameter extrapolation to thermodynamic limit for a system with two holes 
i.e., $n\rightarrow 1$ limit at hopping modulation $\delta t=0.2$. 
These plots are for different exchange coupling $J$ and segment 
length $S_l$ as indicated.}
\label{kr_n12limit}
\end{figure}

\begin{figure}
\subfloat[]{\includegraphics[width = .33\textwidth,angle=-90]{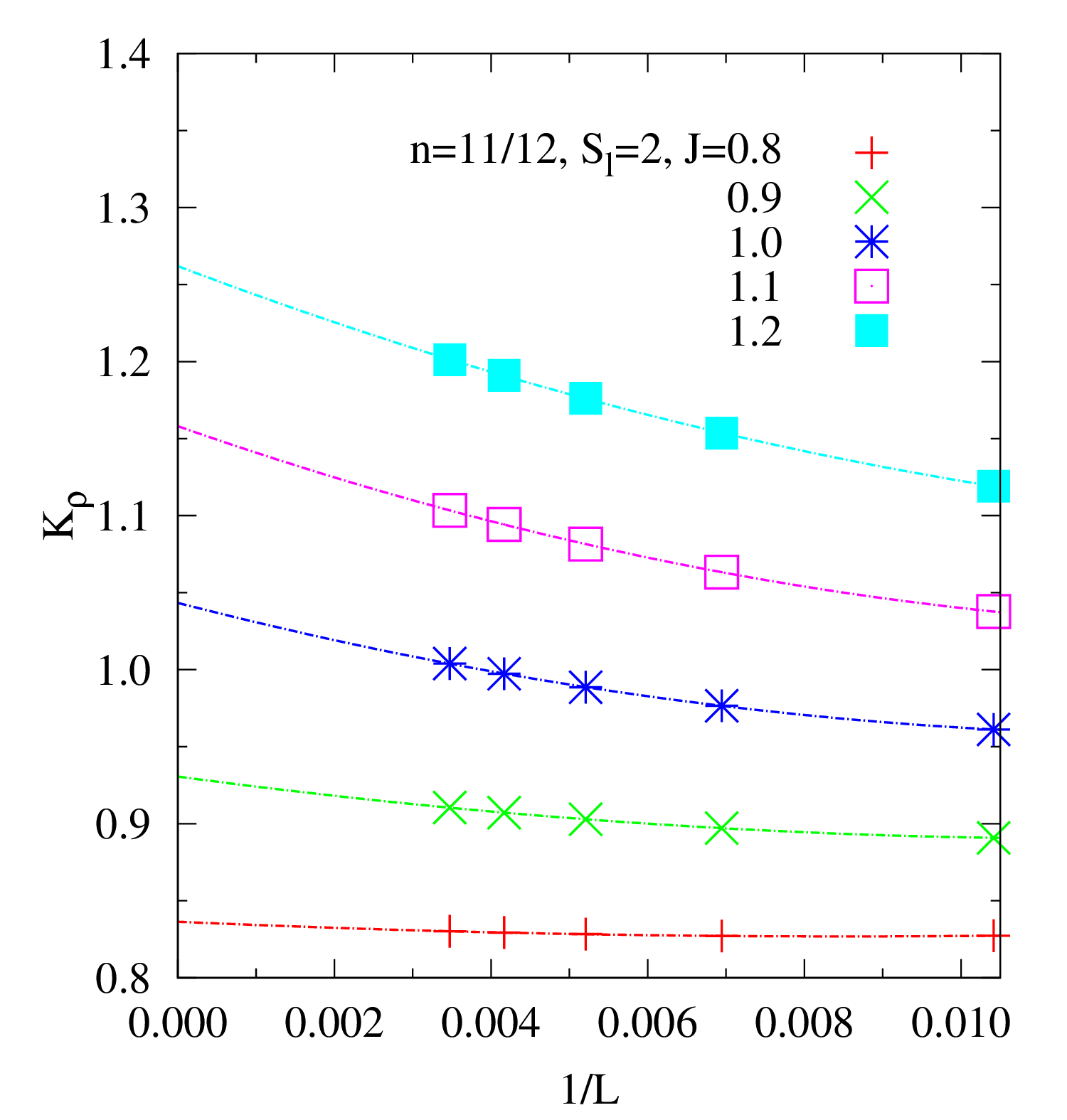}} 
\subfloat[]{\includegraphics[width = .33\textwidth,angle=-90]{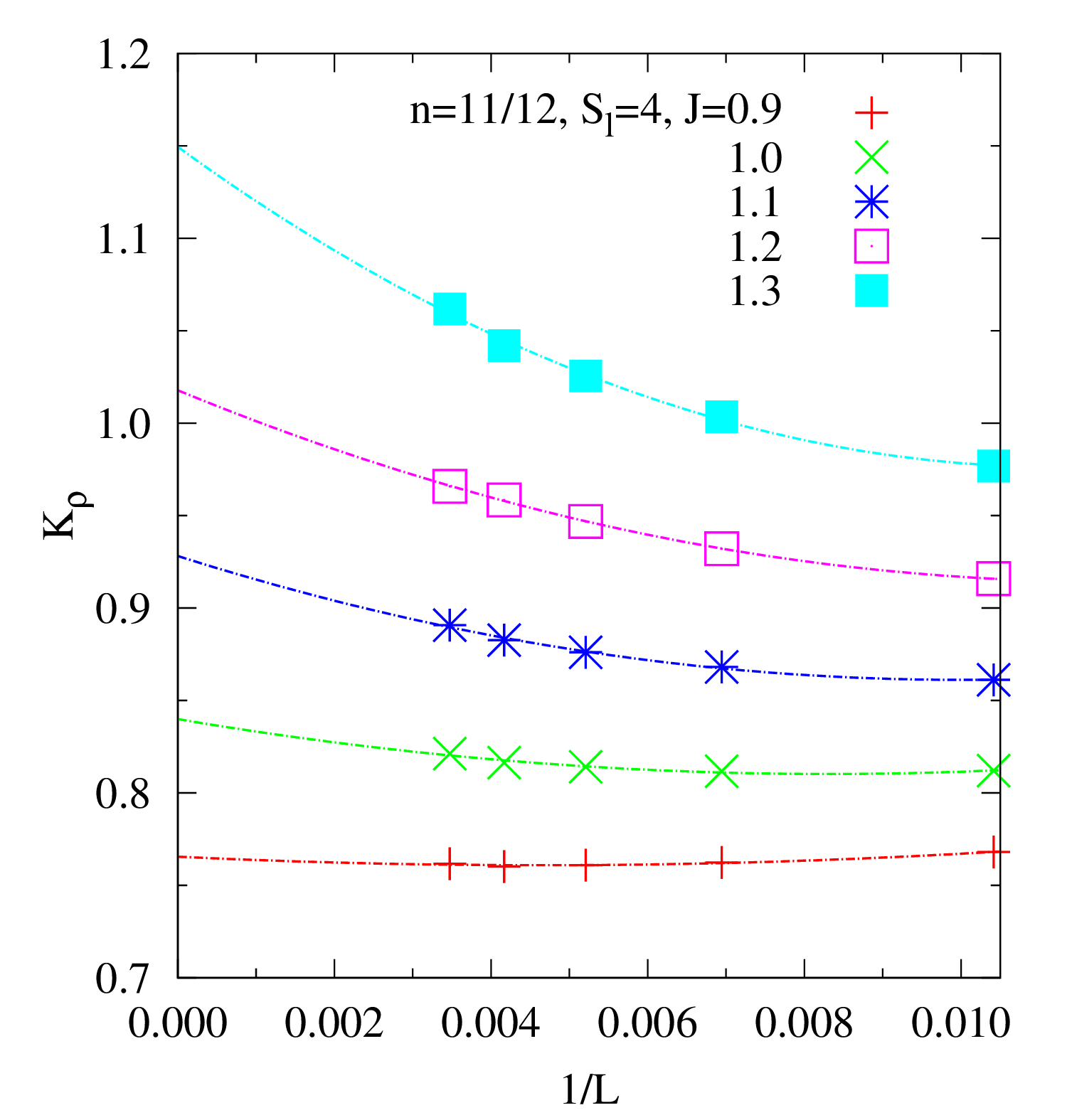}}
\subfloat[]{\includegraphics[width = .33\textwidth,angle=-90]{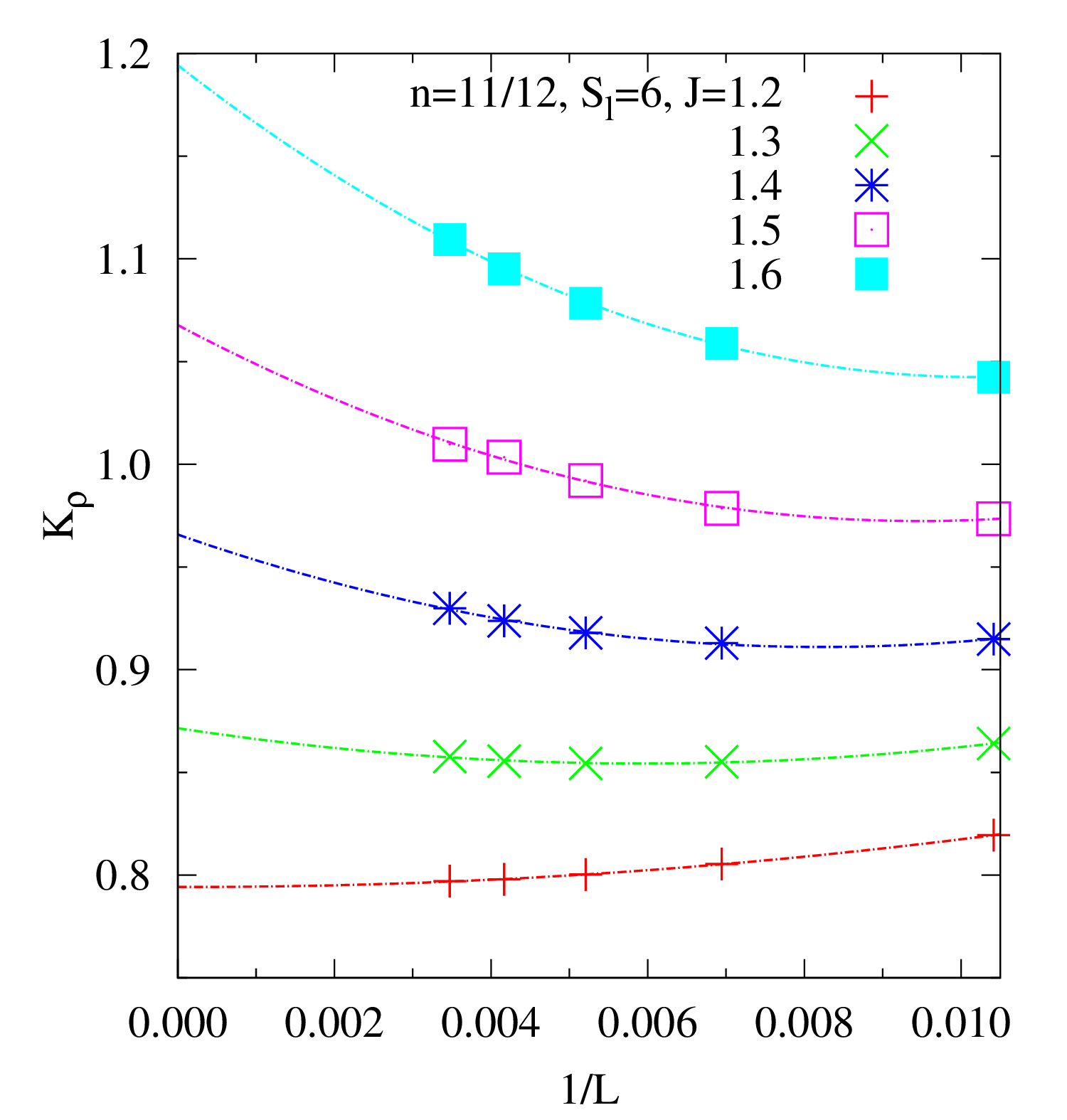}} 
\caption{(Color online) The Luttinger parameter extrapolation to thermodynamic limit for finite density
$n=11/12$. These plots are for different exchange coupling $J$ and segment 
length $S_l$ as indicated.}
\label{kr_n1112}
\end{figure}

\Fig\ref{kr_n12limit} shows the finite size extrapolation to thermodynamic limit for a system 
with two holes i.e., $n\rightarrow 1$ limit at hopping modulation $\delta t=0.2$. We see 
for each segment length $S_l$ the Luttinger parameter $K_\rho$  in thermodynamic limit goes above 
$K_\rho=1$ value where the metallic to superconducting transition happens. The critocal 
value of exchange
coupling $J^c$ at this transition can be as small as $\sim 0.35$ as seen in $S_l=6$ system.
The similar extrapolation of $K_\rho$ at finite density $n=11/12$ and for different $S_l$ system 
are shown in \Fig\ref{kr_n1112}. In this way we have calculated $K_\rho$ in the thermodynamic 
limit for different $n$ and $S_l$ to obtained the $n-J$ phase diagrams shown in the main text.

\begin{figure}
\subfloat[]{\includegraphics[width = .33\textwidth,angle=-90]{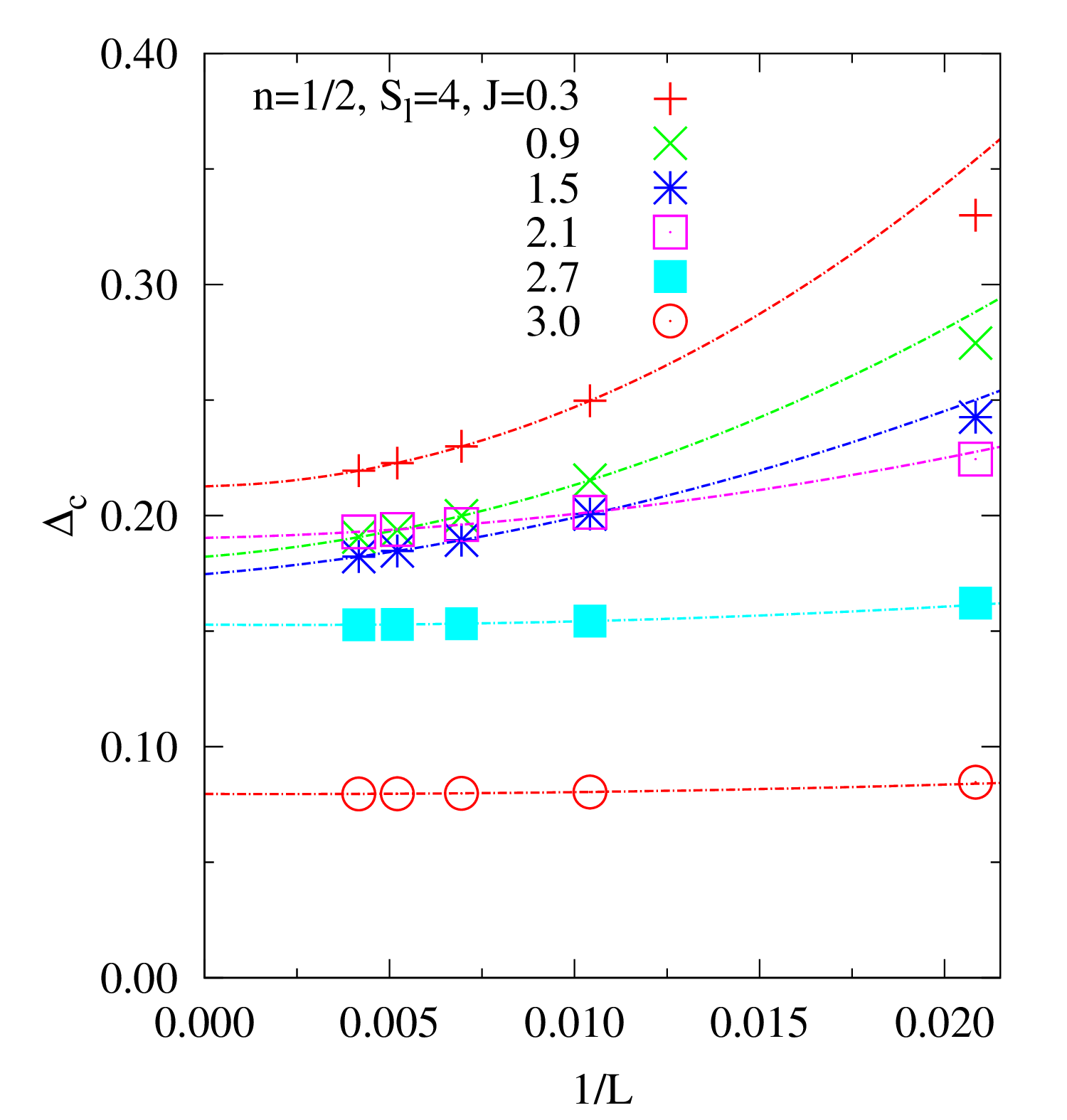}}
\subfloat[]{\includegraphics[width = .33\textwidth,angle=-90]{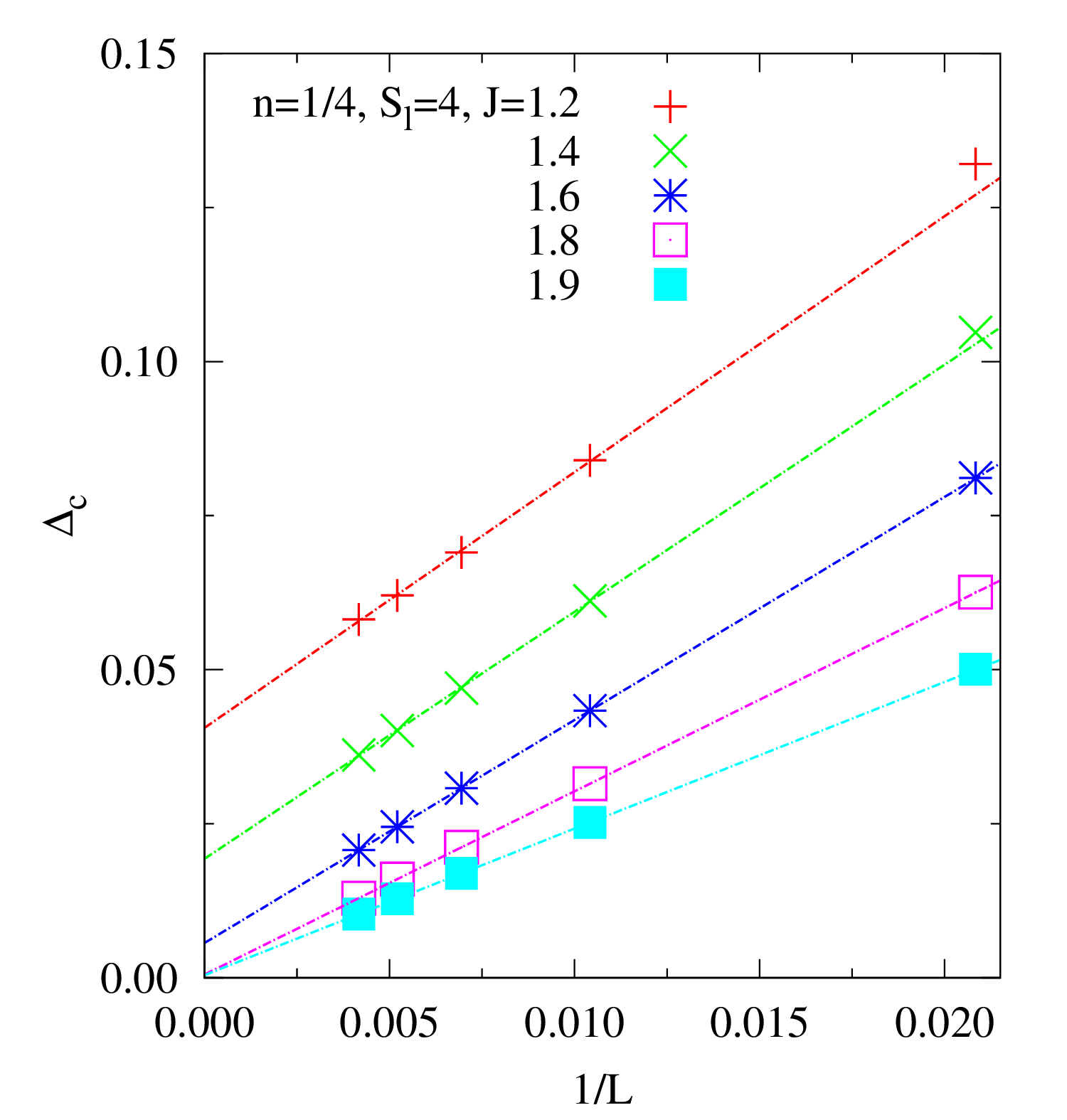}} 
\subfloat[]{\includegraphics[width = .33\textwidth,angle=-90]{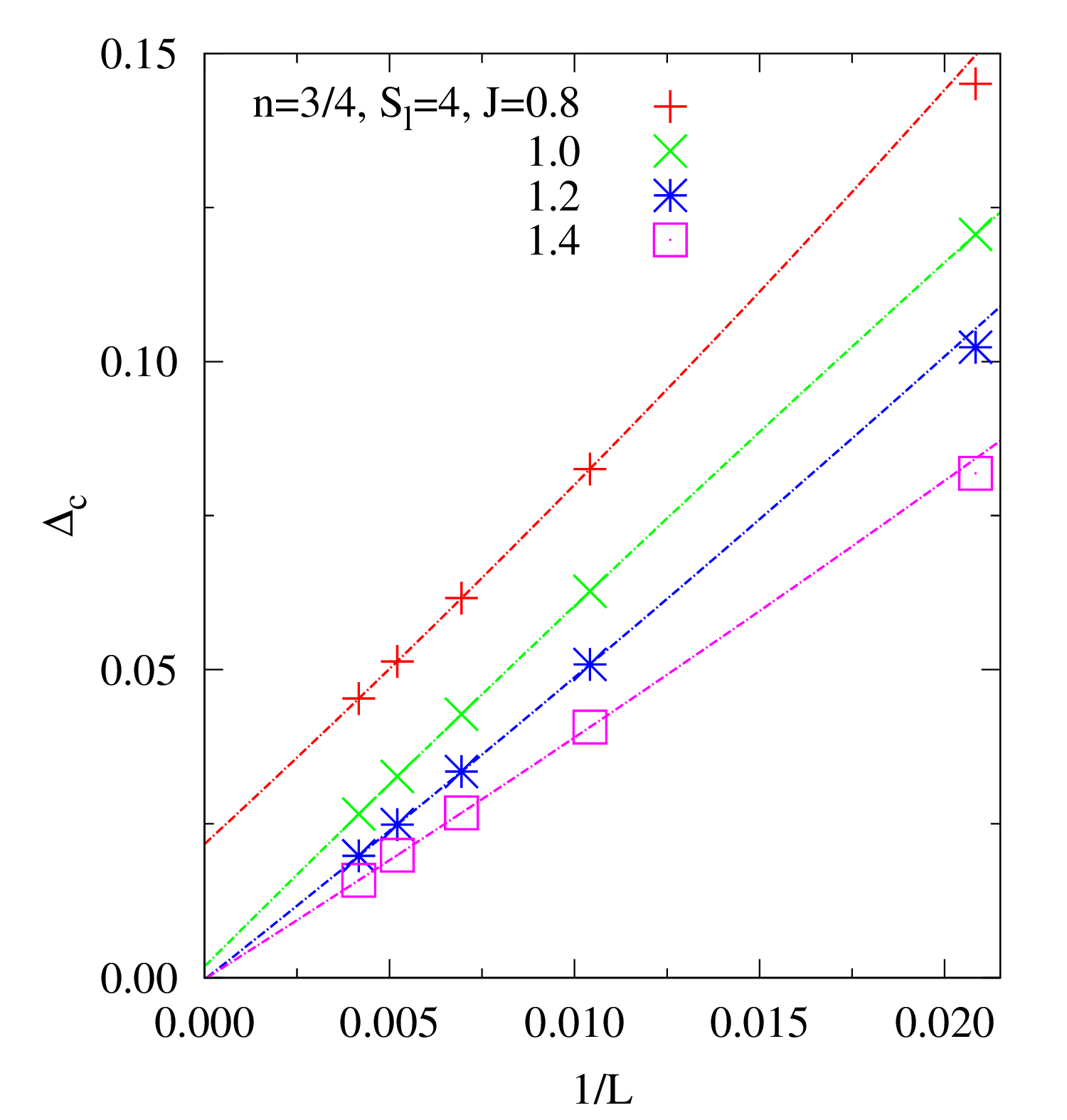}} 
\caption{(Color online) Typical charge gap extrapolation to thermodynamic limit for commensurate densities
for $S_l=4$ system as we increase $J$.}
\label{delc_sl_4}
\end{figure}

\section{Insulating line phases}
Due to the hopping modulation, the electrons tend to form some super structure in each segment
causing insulating behavior at some commensurate fillings for different $S_l$ systems as shown
in $n-J$ phase diagrams in main text. Here we show charge gap $\Delta_c$ 
extrapolation to thermodynamic
limit to show the melting of insulating phase to metallic state as we increase $J$.

As we discussed in detail in the main text that for $S_l=2$ system at $n=1/2$, the competition
between $t-$ and $J-$dimerization causes the melting to metallic states as we increase $J$.  
Here we discuss insulating lines for $S_l=4$ (\Fig 3(b) in main text) system in detail
with the help of charge gap extrapolation.  
At $n=1/2$ we notice that the $S_l=4$
system is insulating for all values of $J$ up to
phase separation. This is because 
two $t-$dimers are formed in each segment for small $J$ and accommodate 
two electrons in bonding states of each $t-$dimer at this commensurate filling
giving a Mott insulator. When we increase $J$, these two electrons already in bonding states
lower the energy by pairing themselves. 
So this insulating phase cannot melt away with increasing $J$ like the $S_l=2$
system at $n=1/2$ (discussed in main text). 
The insulating line phases are captured by calculating the charge gap defined as:
\begin{eqnarray}
\Delta_c=\frac{E(N+2)+E(N-2)-2E(N)}{2}
\end{eqnarray}

\Fig\ref{delc_sl_4}(a) shows the extrapolation of charge gap to thermodynamic limit 
at $n=1/2$ for $S_l=4$ system. The charge gap remains 
finite for all values of $J$ upto phase separation. 
However the melting of insulating states is possible at $n=1/4$ and $3/4$.
At small $J$, one electron per segment (i.e., $n=1/4$) occupies the lowest energy state formed
by superposition of two bonding states already formed by $t-$dimerization. Also for 
$n=3/4$, one electron sits on a $t-$dimer formed on middle two sites 
and other two electrons occupy the ends sites of each segment. As we increase 
$J$, the tendency of $J-$dimerization destroys both the insulating states 
at $n=1/4$ and $3/4$. \Fig\ref{delc_sl_4}(b) and (c) show that the charge gap vanishes as we increase $J$.
It is intuitive that odd number of electrons 
in a segment is not suitable for effective pairings and lead to melting of insulating states as we 
increase $J$. Even number of electrons on the other hand can form an effective 
pairing which prevents the melting of insulating states as shown for $n=1/2$ in $S_l=4$ system. 

\Fig 3(c) of main text depicts the insulating lines at different fillings for $S_l=6$ system. At $n=1/2$, three electrons occupy bonding states in three $t-$dimers formed in each segment for small $J$ leading to a Mott insulator. Again like $S_l=4$ system at $n=1/4$ and $3/4$, the odd number of electrons in each segment can't make effective pair. So with increasing $J$ we have melting of insulating state  at $n=1/2$ due to the competition of $t-$ and $J-$dimerization. The similar argument holds for $n=1/6$ and $5/6$. But with even number of electrons at $n=2/6$ and $4/6$ form effective pairs as discussed in $S_l=4$ system. So the system continues to be insulating upto phase separation. 

\begin{figure}
\begin{center}
\includegraphics[width =.45\columnwidth,angle=-90]{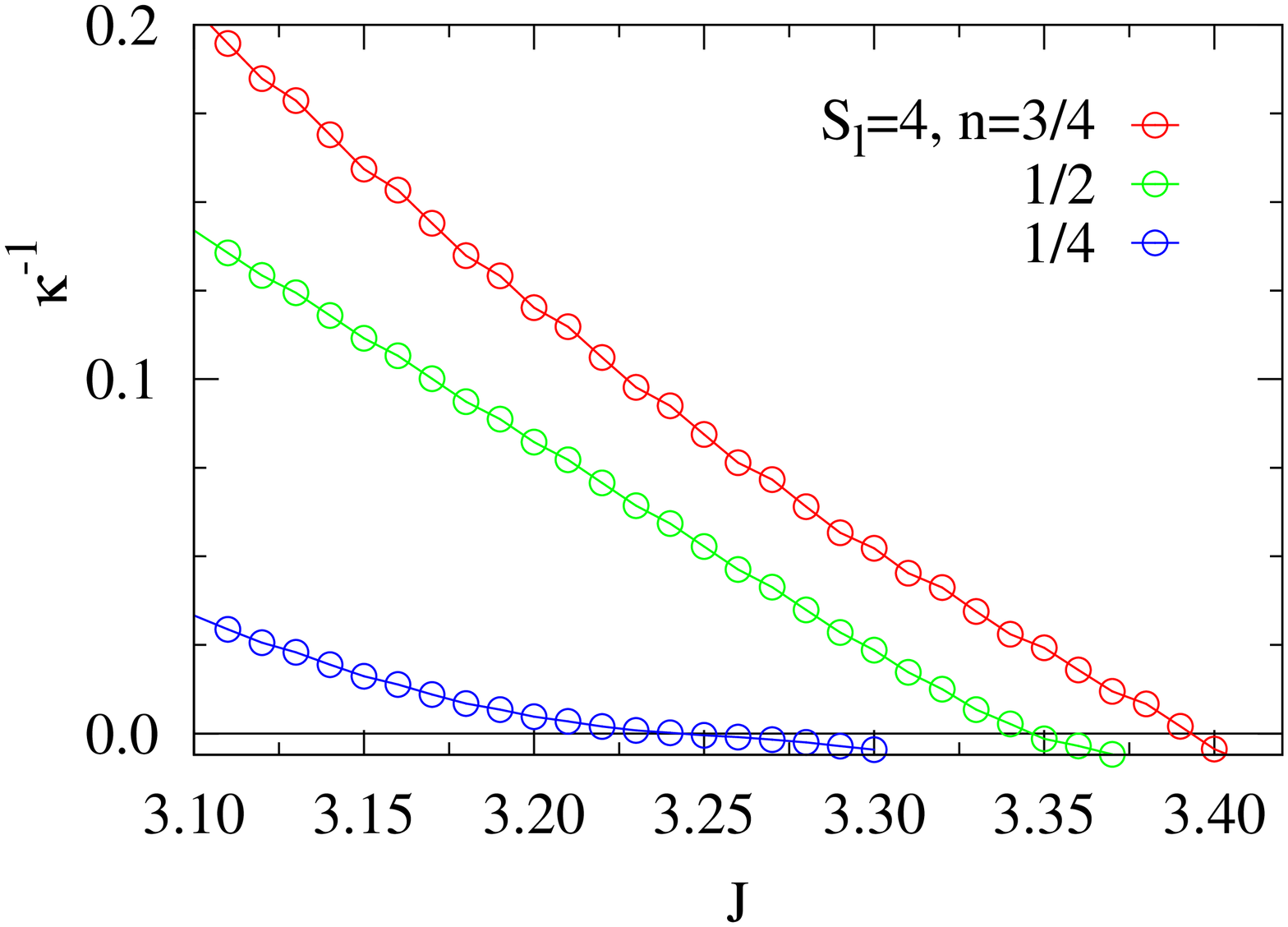} 
\end{center}
\caption{(Color online) Inverse compressibility $\chi^{-1}$ as a function of $J$ at different electron
density for $S_l=4$ system}
\label{phase_sep}
\end{figure} 

Finally in the phase separation the interaction between electrons becomes very large and the electrons
forms antiferromagnetic domains such that the system separates into particle- and hole-rich region. 
We determine the phase separation boundary by calculating inverse compressibility defined as:
\begin{eqnarray}
\chi^{-1}=\frac{N^2}{4}(E(N+2)+E(N-2)-2E(N)) 
\end{eqnarray}
After extrapolating energy to thermodynamic limit, we calculate $\chi^{-1}$ as a function of $J$ and obtain 
the phase separation boundary. \Fig\ref{phase_sep} shows typical plot of $\chi^{-1}$ as function of $J$ at $n=1/4, 1/2$ and 
$3/4$ for $S_l=4$ system. We consider the crossing of $\chi^{-1}$ at zero 
line as onset of phase separation.